\documentclass[iop]{emulateapj}
\usepackage{hyperref}
\usepackage[english]{babel}
\usepackage{amsmath}
\usepackage{natbib}
\usepackage{mathrsfs}
\usepackage{comment}
\shortauthors {B\'{e}ky \& Kocsis}
\shorttitle {Stellar Transits in Active Galactic Nuclei}

\begin {document}

\title {Stellar Transits in Active Galactic Nuclei}

\author {Bence B\'{e}ky, Bence Kocsis\altaffilmark{1}}
\affil {Harvard--Smithsonian Center for Astrophysics, 60 Garden St, Cambridge, MA 02138, USA}
\email {bbeky@cfa.harvard.edu, \\ bkocsis@cfa.harvard.edu}

\altaffiltext {1} {Einstein fellow}

\begin {abstract}
Supermassive black holes (SMBH) are typically surrounded by a dense stellar population in galactic nuclei.
Stars crossing the line of site in active galactic nuclei (AGN)
produce a characteristic transit lightcurve, just like extrasolar planets do when they transit their host star.
We examine the possibility of finding such AGN transits in deep optical, UV, and X-ray surveys.
We calculate transit lightcurves using the Novikov--Thorne thin accretion disk model,
including general relatistic effects.
Based on the expected properties of stellar cusps,
we find that around $10^6$ solar mass SMBHs, transits of red giants are most common for stars on close orbits
with transit durations of a few weeks and orbital periods of a few years.
We find that detecting AGN transits requires repeated observations of thousands of low mass AGNs
to 1\% photometric accuracy in optical, or $\sim10\%$ in UV bands or soft X-ray.
It may be possible to identify stellar transits in the Pan-STARRS and LSST optical and the eROSITA X-ray surveys.
Such observations could be used to constrain black hole mass, spin, inclination and accretion rate.
Transit rates and durations could give valuable information on the circumnuclear stellar clusters as well.
Transit lightcurves could be used to image accretion disks with unprecedented resolution,
allowing to resolve the SMBH silhouette in distant AGNs.
\end {abstract}


\section {Introduction}
\label {sec:introduction}

Photometric transits have been successfully used to identify and characterize transiting extrasolar planets since the discovery of the first one, HD 209458b \citep {2000ApJ...529L..45C, 2000ApJ...529L..41H}. Transit shape is a telltale of planetary, orbital, and stellar parameters.
Moreover, the apparent time-dependent redshift of the star due to the planet covering its receding or approaching side during transit can reveal the projected angle between the planetary orbital axis and the stellar rotational axis \citep[Rossiter--McLaughlin effect,][] {1924ApJ....60...15R,1924ApJ....60...22M}. These methods show that transits are very powerful in probing planetary systems.

Similarly, active galactic nuclei (AGN) accretion disks can be probed by observing occultations in X-ray due to broad line region clouds: optically thick clouds orbiting the supermassive black hole (SMBH) in the region from where broad emission lines of the AGN are thought to originate. These occultations have a large covering factor of $\sim0.1$ to $1$ \citep [see e.g.] [] {1998ApJ...501L..29M, 2008A&A...483..161T, 2009ApJ...695..781B, 2010A&A...517A..47M, 2007ApJ...659L.111R, 2009MNRAS.393L...1R, 2009ApJ...696..160R, 2011MNRAS.410.1027R}. \citet {2011MNRAS.417..178R} pointed out that analogously to the Rossiter--McLaughlin effect, temporally resolved spectroscopic observations of an eclipse could be used to probe the apparent temperature structure of the accretion disk and the origin of the iron K$\alpha$ emission line, allowing to constrain the black hole spin and inclination.

In this paper, we examine the possibility of stars on close orbits transiting their host AGN.
There are multiple coincidences that make it possible to detect such events:
the radius of large stars is comparable to the characteristic size of an accretion disk around a $10^6\;M_\sun$ SMBH, resulting in transits deep enough to detect.
The orbital period is a few years in the innermost regions where the stellar number density is highest,
short enough for repeated observations.
Finally, the typical transit duration for these innermost stars is an hour to a few weeks (depending on the stellar population and the observing band), which is longer than the typical cadence required to observe AGNs, but still accessible on human timescales.

In order to produce a detectable signature in the lightcurve, the transiting object has to be a main sequence O or B star, a Wolf--Rayet (WR) star, an AGB star with a surrounding dust cloud (OH/IR star), a young supermassive star, or an evolved giant. Late type main sequence stars are too small to cause a transit detectable from the ground in an AGN with black hole mass $\gtrsim 10^5\;M_\sun$, unless they are ``bloated'' by irradiation of the AGN \citep{2012ApJ...749L..31L}.

High photometric and temporal resolution observations of AGN stellar transits have the potential to map the accretion disk structure with an unprecedented accuracy. Stars are optically thick in all electromagnetic bands, and unlike broad line region clouds \citep[e.g.][]{2010A&A...517A..47M}, they have a simple spherical geometry, making it easier to interpret lightcurves and to reconstruct the image of the accretion disk along the transit chord (projected stellar trajectory). Furthermore, such transits offer unique observations of individual stars in distant galaxies. Detections of multiple transits could reveal valuable information on the number density of stellar cusp central regions.

In this paper, we calculate transit depths, durations, periods, rates, and instantaneous probabilities based on physical models of nuclear stellar clusters. We also present simulated transit lightcurves and transit depth maps in multiple frequency bands. The shape of the lightcurve depends on the observing wavelength, the mass and spin of the SMBH, the accretion rate, the inclination of the accretion disk, the projected position of the transit chord, and the orbital velocity of the transiting object. Therefore observations with sufficient photometric accuracy and time resolution allow us to constrain these parameters, and to test the employed accretion and general relativity models.

In Section~\ref {sec:cusps},
we review observations of large stars closely orbiting Sgr A*, the SMBH at the center of the Galaxy (\textsection\ref{sec:populations}),
and summarize theoretical models of semi-major axis distribution and mass segregation (\textsection\ref{sec:density}).
We set up models of the stellar population and radial distribution (\textsection\ref{sec:models}),
and determine the minimum (\textsection\ref{sec:rmin}) and maximum (\textsection\ref{sec:rmax}) orbital radii for transits.
We present accretion disk thermal emission models and ray-tracing simulations in Section~\ref{sec:observables}, showing
transit spectra (\textsection\ref{sec:spectra}),
transit depth maps (\textsection\ref{sec:transitmaps})
and lightcurves (\textsection\ref{sec:lightcurves}).
We calculate transit probabilities in Section~\ref{sec:transits}.
In Section~\ref {sec:observability}, we determine the local density of AGNs of interest (\textsection\ref{sec:agndensity}),
and study the feasibility of transit detections with optical (\textsection\ref{sec:optical}, \textsection\ref{sec:kepler})
and X-ray instruments (\textsection\ref{sec:roentgen}).
The most important simplifying assumptions, caveats, and implications are discussed in Section~\ref{sec:discussion}.
Finally, we conclude our findings in Section~\ref{sec:conclusions}.

\section {Nuclear stellar clusters}
\label {sec:cusps}

In this section we review observations and models of stellar distribution in galactic nuclei.

Let $M_\mathrm{SMBH}$ denote the mass of the SMBH, and define $M_6 = M_\mathrm{SMBH}/(10^6\;M_\sun)$ and the gravitational radius $R_\mathrm g = GM_\mathrm{SMBH}/c^2$. Then
\begin {align}
\label {eq:gravrad}
R_\mathrm g &= 4.8\times10^{-5} \;\mathrm{mpc} \; M_6 = 2.1 \; R_\sun \; M_6\,,
\end {align}
where $1\;\mathrm{mpc}=10^{-3}\;\mathrm{pc}=206\;\mathrm{AU}$.
For non-spinning black holes, the Schwarzschild radius is $R_\mathrm S=2R_\mathrm g$, and the radius of the innermost stable circular orbit is $R_\mathrm{ISCO}=6R_\mathrm g$. For maximally spinning black holes, $R_\mathrm S = R_\mathrm g$, and $R_\mathrm{ISCO} = R_\mathrm g$ for prograde orbits in the equatorial plane.

\subsection {Stellar populations}
\label {sec:populations}

Many galaxies host a dense stellar cusp in their nucleus. \citet {2008ApJ...678..116S} find that galaxies with a massive nuclear cluster are more likely to be active. Stars captured and transported inwards by the accretion disk may serve to fuel the AGN \citep{2005ApJ...619...30M}.

In the Galaxy, \citet {2005ApJ...628..246E} report the orbital parameters of six B type main sequence stars orbiting the central SMBH on highly eccentric orbits with semi-major axes less than $16 \;\mathrm{mpc} \sim 10^5\;R_\mathrm g$. These stars may be remnants of stellar binaries tidally disrupted by the SMBH, as first proposed by \citet {1988Natur.331..687H} \citep[see also][] {2003ApJ...599.1129Y}. Candidates for the binary counterparts ejected with high velocity have been identified by \citet {2009ApJ...690.1639B}.

\citet {2009ApJ...697.1741B} find more than a hundred O and WR stars further from the Galactic Center, within $1\;\mathrm{pc} = 5\times10^6\;R_\mathrm g$ projected distance. These stars could have formed in a massive self-gravitating gaseous disk \citep[e.g.][]{2007MNRAS.374..515L,2009MNRAS.394..191H}, or formed further away and captured in close orbits by the \citet {1988Natur.331..687H} mechanism or by a cluster of stellar mass black holes \citep {2004ApJ...606L..21A}. Young stars could also be delivered to this region by an infalling globular cluster \citep{1975ApJ...196..407T,2001ApJ...563...34M}.

Most main sequence stars in the vicinity of a SMBH eventually evolve into red giants or supergiants, our candidates for transiting the AGN. Many such giants have been observed within 1 pc of the Galactic Center \citep[e.g.][]{2010RvMP...82.3121G, 2006ApJ...643.1011P, 2009ApJ...697.1741B, 2010ApJ...708..834B}. The fraction of stars in the giant phase within a stellar population depends strongly on its initial mass function (IMF) and formation history.

A more exotic possibility is the formation of $\sim 10^5\;M_\sun$ supermassive stars in the accretion disk as suggested by \citet {2004ApJ...608..108G}. Such a star would form outside 1000 $R_\mathrm S$, have a radius of approximately $360\;R_\sun$ in case of solar metallicity, and would radiate at its Eddington limit, being luminous enough to have a detectable optical photometric signature not only when it transits the AGN but also when it is occulted behind it.

Observations of the Galactic nucleus show that the innermost cluster of young stars (so-called S-stars) is isotropically distributed \citep{2010RvMP...82.3121G}, as predicted by theoretical arguments. Even if stars are formed on an orbit coplanar with the accretion disk, \citet {1996NewA....1..149R} provide a relaxation mechanism that could rapidly randomize the orbital orientations. The possible presence of intermediate mass black holes may help catalyze this process \citep{2009ApJ...705..361G}. And even if stars cluster in disks, this coherent behaviour averages out when observing multiple galactic cores as long as the stellar disks and the accretion disks have independent orientations. Therefore we assume an isotropic distribution of stellar orbits in the cluster for the purpose of our probability estimates.

\subsection {Density profile}
\label {sec:density}

We assume circular orbits for simplicity, and denote orbital radius by $r$. A star on an eccentric orbit with the same semi-major axis would produce a transit of the same depth, with a transit probability and transit length within a factor of order unity.

\citet {1976ApJ...209..214B} showed that the equilibrium spatial number density distribution of a stellar cluster around SMBH is proportional to $r^{-\alpha}$ with $\alpha = 1.75$ if the stars in the cluster have the same mass. Analytical and numerical investigations of multiple mass populations show that the distribution for each mass bin is still likely to follow a power law. The value of $\alpha$ is predicted to be smaller (down to $\approx1.3$) for lower mass stars \citep {1977ApJ...216..883B, 2006ApJ...649...91F, 2006ApJ...645.1152H}. For massive stars representing a small mass fraction in the stellar cluster, $\alpha$ can be between 2 and $2.75$ \citep {2009ApJ...697.1861A}, or as large as 3 \citep {2009ApJ...698L..64K}. Observations of the Galaxy by \citet {2010ApJ...708..834B} show that WR/O stars from a distance of 30 mpc out to 600 mpc form an interesting anisotropic structure called the clockwise disk in the Galacit nucleus. These stars are distributed with a density exponent $\alpha=2.4\pm0.2$, while the B star population from 30 mpc to 1 pc exhibits $\alpha=2.5\pm0.2$. Note, however, that the age of main sequence O stars and WR stars is less than the two-body relaxation time $\sim0.1$--$1\;\mathrm{Gyr}$ at $r\sim100\;\mathrm{mpc}$ \citep {2009MNRAS.395.2127O}, therefore they are not expected to represent the equilibrium state. The observed mass distribution of solar mass stars in the Galactic nucleus is fit by a broken power law with $\alpha=1.2$ and $1.75$ inside and outside of $0.22\;\mathrm{pc}$, respectively \citep{2007A&A...469..125S}.

To estimate the total number of stars, we first define the radius of influence $r_\mathrm i$ as the radius of the sphere centered on the SMBH containing a total mass of $2M_\mathrm{SMBH}$ in stars and stellar remnants. In case of a singular isothermal sphere ($\alpha=2$), this equals to ${GM_\mathrm{SMBH}}/{\sigma^2}$, where $\sigma$ is the velocity dispersion of stars further than $r_\mathrm i$ \citep {2004cbhg.symp..263M}. In order to get an estimate of the number of stars, we set $r_\mathrm i = {GM_\mathrm{SMBH}}/{\sigma^2}$, independently of $\alpha$.

Using the $M$--$\sigma$ relation now allows us to express the radius of influence as a function the supermassive black hole mass only, in the form
\begin {equation}
\label {eq:radiusofinfluence}
r_\mathrm i = r_0 \; M_6^\gamma\,.
\end {equation}

Here $r_0$ and $\gamma$ depend on the coefficients of the $M$--$\sigma$ relation. In particular, $\gamma=1-2/\beta$, where $\beta$ is the slope of $\log M$--$\log \sigma$, as defined by e.g.~\citet {2002ApJ...574..740T}. For example, the best fit of \citet {2002ApJ...574..740T} (with $\beta=4.02$) results in $r_0=1.234\;\mathrm{pc}$ and $\gamma=0.50$, and the best fit of \citet {2005SSRv..116..523F} (with $\beta=4.86$) results in $r_0=0.881\;\mathrm{pc}$ and $\gamma=0.59$. For our numerical results, we adopt the best fit values of \citet {2009ApJ...698..198G} (with $\beta=4.24$): $r_0=1.075\;\mathrm{pc}$ and $\gamma=0.53$ (but use 1 pc to normalize $r_0$ in our parametric expressions).

Now let us consider the stellar population in the vicinity of the SMBH. We assume that this population contains a species of stars with mass $M_\star$, having a radius $R_\star$ large enough to produce detectable transits. We also assume that the number density of these stars follows the power law $r^{-\alpha}$. Let $b$ denote the mass fraction of these large stars within the sphere of influence relative to the total mass of all stars and stellar remnants.

Typically the inner cutoff radius $r_\mathrm{min}$ for the stellar distribution is much smaller than the radius of influence (see \textsection~\ref {sec:rmin} for numerical justification). Under this assumption, the spatial number density of the stars in question as a function of orbital radius is
\begin {equation}
\label {eq:density}
n(r) = b \frac {3-\alpha}{2\pi} \frac{M_\mathrm{SMBH}}{M_\star} \frac{r^{-\alpha}}{r_i^{3-\alpha}}\,.
\end {equation}

Note that this argument has two weaknesses: first, the $M$--$\sigma$ relation has a relatively large scatter. For a given SMBH mass, the intrinsic scatter of the bulge velocity dispersion is $\sim0.3$ dex \citep{2009ApJ...698..198G}. Second, we applied results for the isothermal sphere to power law distributions with different exponents. This limits the accuracy of the transit rate estimates presented in Section~\ref{sec:transits}.

\subsection {Adopted models}
\label {sec:models}

\begin{deluxetable*}{ccccccccccccccc}
\tabletypesize{\scriptsize}
\tablecaption{Orbital radius and period limits for AGN transits for the adopted stellar population models \label{tab:rlimits} ($b=0.01$ for each model)}
\tablehead{ $M_\mathrm{SMBH}$ & model & $R_\star$ & $R_\star$ & $M_\star$ & $\alpha$ & $r_\mathrm {tid}$ & $r_\mathrm{coll}$ & $r_\mathrm{min}$ & $r_\mathrm{lens}$ & $r_\mathrm i$ & $r_\mathrm{max}$ & $T_\mathrm{orb}\left(r_\mathrm{min}\right)$ & $T_\mathrm{orb}\left(r_\mathrm{max}\right)$ \\
$(M_\sun)$ && $(R_\sun)$ & $(R_\mathrm g)$ & $(M_\sun)$ && (mpc) & (mpc) & (mpc) & (mpc) & (mpc) & (mpc) & (yr) & (yr) }
\startdata
$10^5$ & O stars    &  11 & 51.8  & 30  & 2.5  & 0.006 & 0.16 & 0.16 &      11 &    320 &     11 & 0.57 &      340 \\
$10^5$ & red giants & 110 & 518   & 1.5 & 1.75 & 0.16  & 0.54 & 0.54 & 22\,000 &    320 &    320 & 3.7  &  53\,000 \\
$10^6$ & O stars    &  11 & 5.18  & 30  & 2.5  & 0.013 & 0.40 & 0.40 &      11 & 1\,075 &     11 & 0.75 &      110 \\
$10^6$ & red giants & 110 & 51.8  & 1.5 & 1.75 & 0.35  & 1.3  & 1.3  & 22\,000 & 1\,075 & 1\,075 & 4.2  & 100\,000 \\
$10^7$ & O stars    &  11 & 0.518 & 30  & 2.5  & 0.028 & 1.0  & 1.0  &      11 & 3\,600 &     11 & 1.0  &       34 \\
$10^7$ & red giants & 110 & 5.18  & 1.5 & 1.75 & 0.76  & 3.0  & 3.0  & 22\,000 & 3\,600 & 3\,600 & 4.9  & 210\,000
\enddata
\end{deluxetable*}

We consider two simple models for the transiting stellar populations around AGNs, summarized in Table~\ref{tab:rlimits}.

First, we assume a young stellar population, motivated by the observations of a young population with an extremely top-heavy initial mass function of mean stellar mass $\approx 30\;M_\sun$ following a density profile with $\alpha \approx 2.5$ in the central parsec of the Galaxy \citep{2010ApJ...708..834B}. Thus in our model, we assume that a fraction of the stars are O type, with stellar mass $M_\star= 30\;M_\sun$, radius $R_\star=11\;R_\sun$, and $\alpha=2.5$. This exponent is consistent with observations and theoretical predictions reviewed in Section~\ref{sec:density}. We assume that these O stars constitute a mass fraction $b=0.01$ of the population. This is consistent with the estimated total mass of WR/O stars in the Galactic Center if we consider that these stars are confined in the center part of the sphere of influence. However, note that this mass fraction depends sensitively on recent star formation rate and initial mass function of stars in the neighborhood of the SMBH.

For the second model, we consider an evolved, relaxed population of stars, and optimistically assume that a $b=0.01$ mass fraction of them are giants (or main sequence stars otherwise enlarged, like ``bloated'' or surrounded by a dust cloud), which are large enough to produce detectable AGN transits. For these giants, we assume $R_\star = 110\;R_\sun$, $M_\star=1.5\;M_\sun$, and a Bahcall--Wolf equilibrium value of $\alpha=1.75$ (see Section~\ref{sec:density}).

Note that depending on the star formation history, O stars and red giants might coexist in the cusp, in which case their contributions to transits add up.

\subsection {Minimum orbital radius}
\label {sec:rmin}

The inner orbital radius cutoff of the stellar distribution, $r_\mathrm{min}$, is set by gravitational wave inspiral, and tidal and collisional disruption. While gravitational wave inspiral is the limiting factor for compact objects \citep{1964PhRv..136.1224P}, tidal or collisional disruption sets a tighter constraint for stars.

The tidal disruption radius is
\begin {align}
r_\mathrm t &= \left( \eta^2 \frac {M_\mathrm {SMBH}} {M_\star} \right) ^{\frac13} R_\star\nonumber\\
&= 0.013\;\mathrm{mpc} \; M_6^{\frac13} \left( \frac {M_\star} {30\;M_\sun} \right) ^{-\frac13} \left( \frac {R_\star} {11\;R_\sun} \right)\,,
\end {align}
where $\eta$ ranges from $0.8$ to $3.1$ depending on the compressibility and polytropic mass distribution index of the star \citep {1995MNRAS.275..498D}. We adopt the moderate value $\eta=2$. The tidal disruption radius is given here normalized for a massive main sequence star.

Collisional disruption might be responsible for the depletion of luminous giant stars in the inner 80 mpc of the Galactic Center \citep {1999ApJ...527..835A}. Following \citet {2007MNRAS.378..129H}, one can write that the rate at which collisions decrease stellar density is
\begin {equation}
\frac{\partial n(r,t)}{\partial t} = - \frac {n^2(r,t) v(r) \Sigma}{N_\mathrm{coll}},
\end {equation}
where $\Sigma = \pi R_\star^2$ is the collisional cross-section, and on average, $N_\mathrm{coll}\approx30$ collisions are required to disrupt a star \citep{2006ApJ...649...91F}. We define the radius limit for collisional disruption as the radius where the stellar density $e$-folds in time $t$ if the initial value is given by Equation \ref {eq:density}:
\begin {equation}
r_\mathrm{coll} = r_0 \left[ \vphantom {\left( \frac{R_\star}{11\;R_\sun}\right)^2}
1.33 \times 10^{-8} \; b (3-\alpha) M_6^{\frac32 - \gamma(3-\alpha)}\right. \times
\end {equation}
\begin {equation*}
\times \left. \left( \frac {t}{10^7\;\mathrm{yr}} \right) \left( \frac {r_0}{1\;\mathrm{pc}} \right)^{-\frac72} \left( \frac {M_\star}{30\;M_\sun} \right)^{-1} \left( \frac{R_\star}{11\;R_\sun}\right)^2 \right]^{\frac2{2\alpha+1}}.
\end {equation*}
We set the timescale to be $t=10^7\;\mathrm{yr}$: this is within an order of magnitude of both the main sequence lifetime of early type stars, and the lifetime of the giant phase for less massive stars. Note that here we only consider collisions within the large species, not with other, much smaller stars, which are less likely.

The minimum orbital radius is
\begin {equation}
r_\mathrm{min} = \max ( r_\mathrm t, r_\mathrm{coll})\,.
\end {equation}
Table~\ref{tab:rlimits} lists the minimum and maximum orbital radii for the two stellar species and three different SMBH masses. We find that in every case, collisions set a tighter constraint than tidal disruption for the potentially transiting stars.

Table~\ref{tab:rlimits} also lists the Keplerian orbital periods for stars at the minimum and maximum radii, which can be calculated as
\begin {align}
\label {eq:torb}
T_\mathrm{orb} &= 3\;\mathrm{yr} \; \left(\frac r {1\;\mathrm{mpc}} \right)^{\frac32} \; M_6^{-\frac12}.
\end {align}
We find that the closest main sequence stars have orbital periods of approximately one year, making repeated transit observations feasible. However, collisions set a larger radius limit for giants, resulting in longer orbits. This inner radius limit depends on the number fraction of giants: smaller $b$ implies less frequent collisions and thus allows closer orbits. However, a smaller value for $b$ would also mean smaller transit probabilities for a given AGN (see Section~\ref{sec:transits} below).

\subsection {Maximum orbital radius}
\label {sec:rmax}

For the maximum orbital radius $r_\mathrm{max}$ of stars capable of producing a transit, we have to consider two factors: gravitational microlensing due to the star, and the validity range of the presumed number density power law.

Gravitational microlensing caused by the transiting star can bend the light rays of the AGN which may significantly distort the transit lightcurve \citep{1986ApJ...304....1P}. This can happen if the transiting object is farther from the AGN than the radius $r_\mathrm{lens}$ at which the Einstein radius equals the apparent angular radius of the transiting object:
\begin {equation}
r_\mathrm{lens} = 11\;\mathrm{mpc} \left(\frac{R_\star}{11\;R_\sun}\right)^2 \left(\frac {M_\star}{30\;M_\sun}\right)^{-1}\,.
\end {equation}
Therefore we restrict our transit probability calculations to the contribution of stars within orbital radius $r_\mathrm{lens}$.

Note that this is a different configuration than a galaxy microlensing a distant quasar, which can also be used to probe the spatial structure of accretion disks around AGNs \citep[e.g.~for the case of Q2237+0305, see][and references therein]{2004ApJ...605...58K}.

The power law distribution discussed above only applies to the stellar populations within the radius of influence from the SMBH, where its gravitational field dominates. In this paper, we do not investigate the distribution of stars outside the sphere of influence, but conservatively ignore their contribution to transit probabilities.

The maximum orbital radius is thus the smaller of the microlensing radius and the radius of influence:
\begin {equation}
r_\mathrm{max} = \min (r_\mathrm{lens}, r_\mathrm i)\,.
\end {equation}
Table~\ref{tab:rlimits} shows that typically microlensing is the limiting factor among main sequence O stars, whereas the radius of influence limits transits of the much less dense red giants. 

\section {Transit observables: spectra and lightcurves}
\label {sec:observables}

Next we derive the AGN spectra and the transit observables.

The AGN luminosity is bounded by the Eddington limit $L_\mathrm{Edd} = 1.3 \times 10^{44} \;\mathrm{erg}\;\mathrm s^{-1}\; M_6$ \citep {1983bhwd.book.....S}. We assume an Eddington ratio of $0.25$ as our fiducial value, following \citep {2006ApJ...648..128K,2011arXiv1111.3574S}. Then the AGN luminosity is
\begin {equation}
\label {eq:ledd}
L_\mathrm{AGN} = 3.6\times10^{43}\;\mathrm{erg}\;\mathrm s^{-1}\; M_6\,.
\end {equation}

\subsection {AGN spectra}
\label {sec:spectra}
We adopt the general relativistic, radiatively efficient, steady-state thin accretion disk model of \citet {1973blho.conf..343N}. This model describes the flux of thermal radiation from the disk as a function of radius in Equation (5.6.14b) \citep [see][for the explicit form of $\mathscr Q$] {1974ApJ...191..499P}. We assume no radiation from within $R_\mathrm{ISCO}$. In addition to the thermal component, AGN spectra typically exhibit emission lines, excess infrared radiation from dust reprocessing UV emission, and a hard X-ray component usually assigned to inverse Compton scattering in a hot corona \citep {1993ApJ...413..507H}. We do not account for these phenomena, but choose our observing bands so that their effect is minimal: an observation window around 200 eV is low enough so that the thermal component dominates over coronal emission, but it is higher than helium Lyman absorption and detector lower energy limits. It is important to note that little is known about the geometry of the corona, and simultaneously observing a stellar transit in hard X-ray might provide feedback to the models.

We follow the accretion disk photosphere model described by \citet {2005ApJ...622L..93M} and \citet {2010ApJ...714..404T}: we assume that the dominant absorption mechanism is the bound-free process, with an opacity that depends on the frequency and temperature. We assume that the temperature and thus the absorption opacity are constant down to an optical depth of one (the ``bottom'') in the photosphere. We can calculate the total flux from the absorption and electron scattering opacities and the photosphere bottom temperature, using Equations (A13--A15) and (A17) of \citet {2010ApJ...714..404T}, but using the relativistic angular frequency given by \citet {1973blho.conf..343N} instead of Keplerian velocity. However, as the flux is known and the temperature is sought for, we have to use a simple iterative process to solve this implicit equation for the temperature. We find that usually electron scattering dominates the total opacity, but in the hottest parts of the accretion disk, absorption takes over. Note that the functional form of the specific flux differs from a blackbody spectrum, because the absorption opacity depends on frequency. For simplicity, we assume that the emerging radiation is isotropic in the frame comoving with the accretion disk.

\begin {figure}
\includegraphics*[width=86mm]{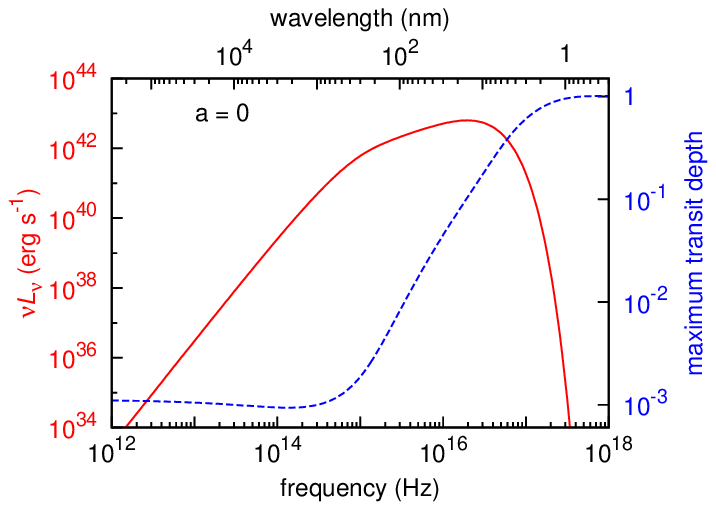}
\includegraphics*[width=86mm]{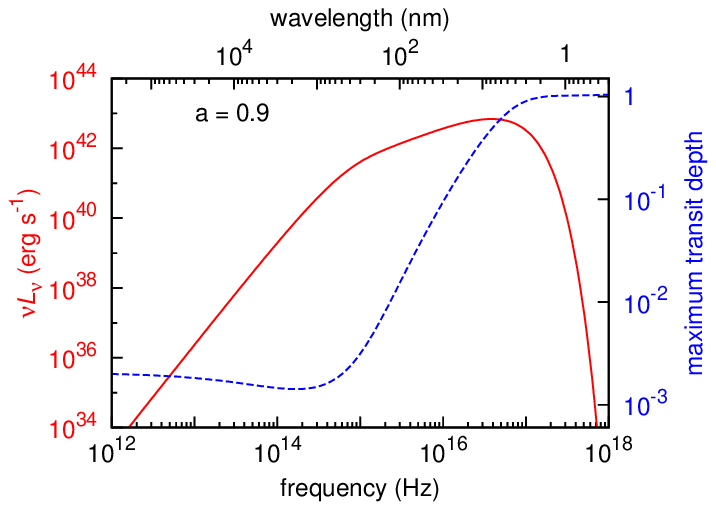}
\caption {Spectrum and transit depth of an accretion disk around a $10^6\;M_\sun$ black hole. Solid red curve: equivalent isotropic luminosity per logarithmic bins of frequency. Dashed blue curve: maximum possible narrow-band transit depth caused by a star with $R_\star=11\;R_\sun$ as a function of source frequency. Top (bottom) panel presents the case of a Schwarzschild (Kerr) BH with spin $a=0$ ($a=0.9$).}
\label {fig:freq}
\end {figure}

Given the specific intensity of the accretion disk as a function of radius and frequency, the observed spectrum is determined by Doppler shift, gravitational redshift, and gravitational lensing. To account for these effects, we apply the transfer function method as described by \citet {1975ApJ...202..788C} and implemented by \citet {1995CoPhC..88..109S}\footnote {available at \href{http://www.tat.physik.uni-tuebingen.de/~speith/publ/photon_transferfct_dble.f}{http://www.tat.physik.uni-tuebingen.de/}\\\href{http://www.tat.physik.uni-tuebingen.de/~speith/publ/photon_transferfct_dble.f}{\textasciitilde{}speith/publ/photon\_transferfct\_dble.f} }. We calculate the radiative efficiency as a function of spin as described by \citet {1983PhT....36j..89S} to convert luminosity to mass accretion rate, which is the input parameter of this code. We fix the inclination angle $\vartheta$ (the angle between the spin axis and the line of sight) at $60^\circ$, so that $\cos\vartheta=0.5$. The calculated spectrum of the accretion disk is shown in Figure~\ref{fig:freq} for a non-spinning black hole (with dimensionless spin $a=0$) in the top panel, and for a highly spinning black hole with a prograde disk with a conservative value $a=0.9$ in the bottom panel. The frequency is in the source rest-frame accounting for gravitational redshift, but a possible cosmological redshift for distant AGNs is not considered. The specific luminosity value displayed here is what an isotropic source would have to have in order to produce the same flux as the AGN does at this specific inclination.

Transit depth is defined as the blocked flux to out-of-transit flux ratio in a given band. Therefore the transit depth is between zero and one: zero if the transiting object does not cover any part of the accretion disk; one if the object completely blocks radiation (in which case it is called an occultation or eclipse). The transit depth varies with frequency and the location of the transiting object in projection. To be able to efficiently calculate transit depths at different positions, we implement a linear approximation to the radius--gravitational redshift grid generated by the above code to determine these values for a light ray parametrized by its projected position far from the AGN. Then we employ high order numerical approximation\footnote {code available from \href{http://www.holoborodko.com/pavel/?page_id=1879}{http://www.holoborodko.com/pavel/}\\\href{http://www.holoborodko.com/pavel/?page_id=1879}{?page\_id=1879}} to integrate over the stellar disk in the projection plane. This gives the blocked specific flux, which we then divide by the total specific flux to obtain the narrow-band transit depth.

In addition to AGN spectra, Figure~\ref{fig:freq} also depicts the maximum possible narrow-band transit depth caused by an early type main sequence star with $R_\star=11\;R_\sun$ as a function of frequency. The maximum depth of a general transit can be smaller if the star does not transit the most luminous part of the projected accretion disk. At high frequencies, the most luminous region is more compact, therefore the transit is deeper. The transit depth becomes constant at infrared frequencies less than the peak frequency of a black body spectrum with temperature of the outer edge of the disk (we set $10^3\;R_\mathrm g$ in the simulations), because here the Rayleigh--Jeans approximation applies to every part of the disk. Larger stars cause deeper transits (see below).

\subsection {Transit maps}
\label {sec:transitmaps}

\begin {figure}
\includegraphics*[width=86mm]{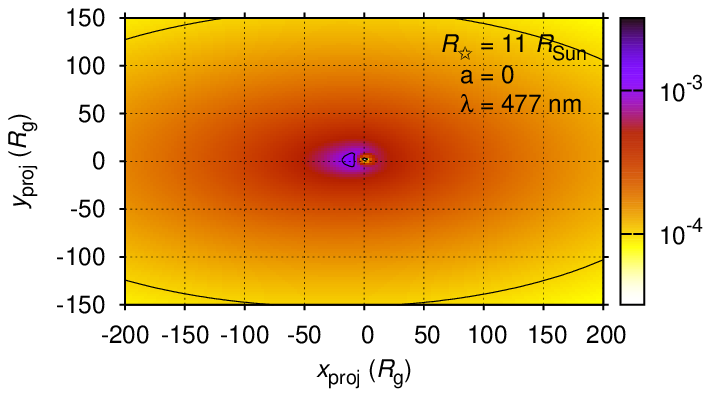}
\includegraphics*[width=86mm]{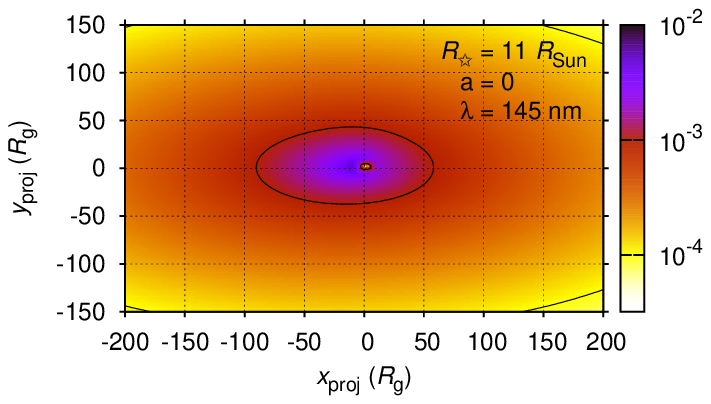}
\includegraphics*[width=86mm]{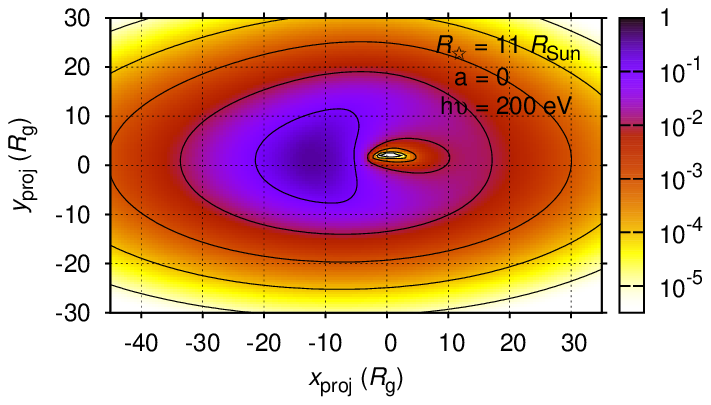}
\caption {The instantaneous narrow-band transit depth as a function of projected position of the transiting object at three different frequencies optical ({\it top panel}), EUV ({\it middle panel}), and soft X-rays ({\it bottom panel}). A transit lightcurve corresponds to the values along the stellar trajectory in the image plane shown. A non-spinning $10^6\;M_\sun$ black hole is at the origin, the observation angle is $60^{\circ}$ relative to the accretion disk, and transit depths are shown for a main sequence O-type star with $R_\star=11\;R_\sun=5.18\;R_\mathrm g$. The top half of the disk image is more severely distorted by gravitational lensing since it is farther from the observer than the black hole. The left side of the disk rotates towards the observer, thus appearing more luminous due to beaming, which results in a deeper transit. The transit is deepest if the projected position of the star crosses the most luminous regions closest to the SMBH. The black hole silhouette and the curved spacetime geometry becomes visible in the X-ray transit map.
}
\label {fig:transitmapostar}
\end {figure}

\begin {figure}
\includegraphics*[width=86mm]{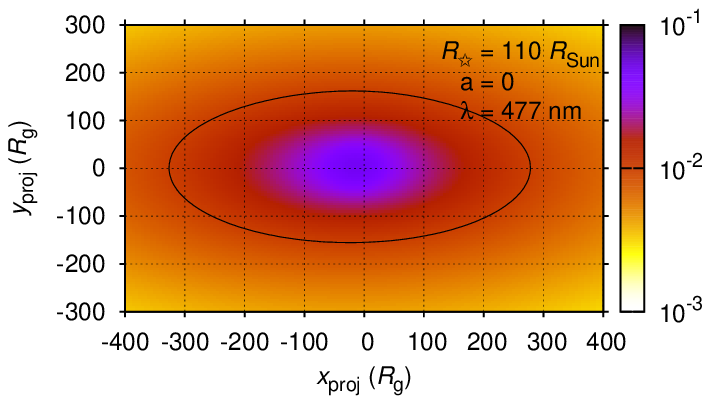}
\includegraphics*[width=86mm]{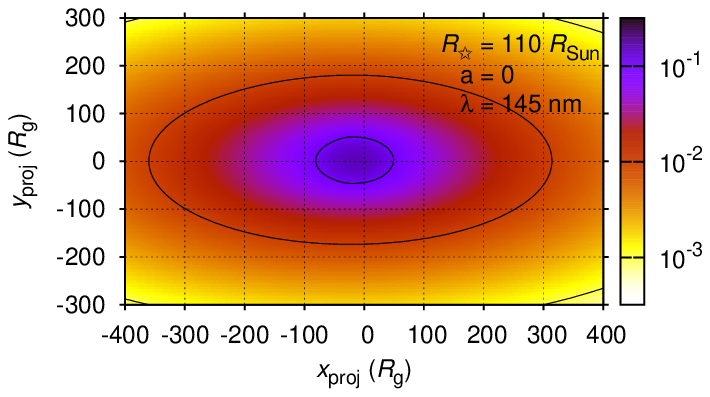}
\includegraphics*[width=86mm]{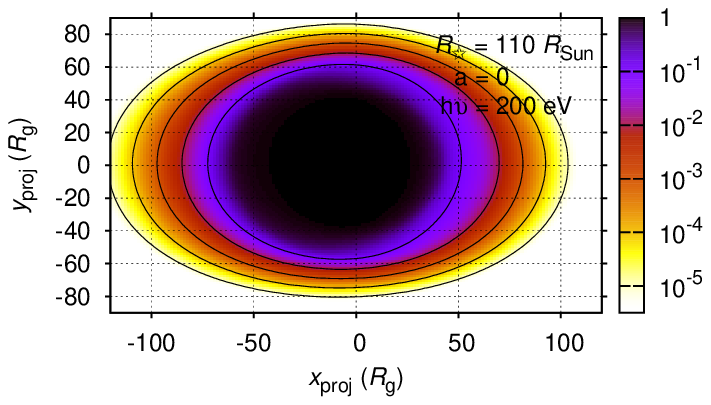}
\caption {Same as Figure~\ref {fig:transitmapostar}, but for a giant star with $R_\star=110\;R_\sun=51.8\;R_\mathrm g$. Here, the star is so large that the transit is much deeper, and the image of the accretion disk is smoothed out (e.g.~the left-right asymmetry of beaming is not apparent).}
\label {fig:transitmapgiant}
\end {figure}

\begin {figure}
\includegraphics*[width=86mm]{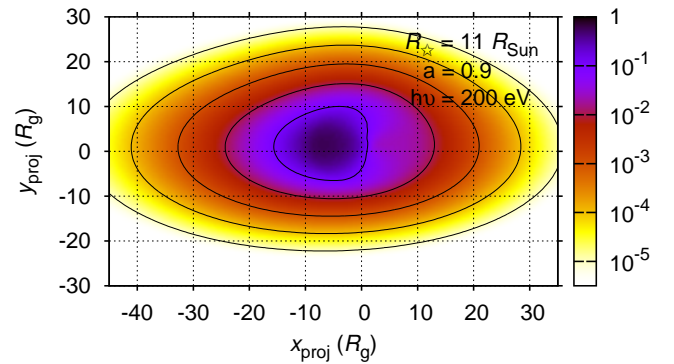}
\caption {Soft X-ray transit depth map for an AGN with a Kerr-BH spin $a=0.9$ for $R_\star=11\;R_\sun$, same as the bottom panel of Figure~\ref {fig:transitmapostar} but for a spinning SMBH. The details of the accretion disk are smoothed since the ISCO radius in this case is smaller than the stellar radius. The accretion disk image is further modified because of the Kerr geometry.}
\label {fig:transitmapspin}
\end {figure}

As stated in Section~\ref {sec:introduction}, AGN transit observations can be used to map distant AGNs along the transit chord with unprecedented resolution. We now elaborate on this point. Figures~\ref{fig:transitmapostar} and \ref{fig:transitmapgiant} show the transit depth as a function the projected position of the transiting object in front of the accretion disk around a non-spinning BH for an early type main sequence star ($R_\star=11\;R_\sun$) and a giant star ($R_\star=110\;R_\sun$), respectively. The three panels in both figures show the transit depth maps for different observing frequency: optical $g$ band (top panel, $\lambda = 477 \;\mathrm{nm}$, $\nu = 6.3 \times10^{14} \;\mathrm{Hz}$), extreme ultraviolet (EUV, middle panel, $\lambda = 145 \;\mathrm{nm} = 1450$ \AA, $\nu = 2.1 \times10^{15} \;\mathrm{Hz}$), and soft X-ray (bottom panel, $h\nu=200 \;\mathrm{eV}$, $\nu = 4.8 \times10^{16} \;\mathrm{Hz}$, $\lambda = 6.2 \;\mathrm{nm}$). The spatial variations in the transit depth maps imply time-variations of the observed AGN brightness as a transiting object moves across the image. A transit lightcurve corresponds to the values along the projected stellar trajectory in Figures~\ref{fig:transitmapostar} and \ref{fig:transitmapgiant}. Conversely, observations of the transit lightcurve can be used to reveal the geometry of the accretion disk along a line in this image.

The maximum possible resolution of such a reconstructed image is set by the size of the transiting object and the spatial variations of the disk brightness. Since the disk temperature increases inwards, the emission at higher frequencies is confined to the innermost regions, implying that transit measurements at higher frequencies can give a higher resolution image (see different panels in Figures~\ref{fig:transitmapostar} and \ref{fig:transitmapgiant}). Transits of smaller objects also yield a higher resolution image. However, transits of smaller objects are less deep and hence more difficult to detect.

The left-right asymmetry in the figure is due to different Doppler shifts for regions of the disk moving towards or away from the observer. The spacetime geometry leaves an imprint on the image, the top half of the disk image is distorted by gravitational lensing close to the BH. Remarkably, the black hole silhouette (i.e., the lack of emission within the ISCO) becomes directly visible in the X-ray image of a transiting O star (see Figure~\ref{fig:transitmapostar} bottom panel).

The transit depth map is also sensitive to the spacetime geometry both directly through gravitational lensing and indirectly through the change in the ISCO radius. Figure~\ref{fig:transitmapspin} shows the soft X-ray transit map of a Novikov--Thorne accretion disk around a Kerr BH with spin $a=0.9$ (c.f.~bottom panel of Figure~\ref{fig:transitmapostar}). In this case, the ISCO is smaller than the radius of the transiting object and the BH silhouette does not appear in the image.

\subsection {Transit lightcurves}
\label {sec:lightcurves}
To get a handle on the characteristic transit duration, let us consider the timescale for the center of a star to cross a disk of radius $R_\mathrm{AGN}$ centered on the most luminous part of the accretion disk image for a given frequency. We choose the value of $R_\mathrm{AGN}$ based on the transit depth map, depending on the desired transit depth. Typically $R_\mathrm{AGN}\sim5$--$1000\;R_\mathrm g$.

The transit duration is
\begin {align}
\label{eq:Deltat}
\Delta t &= 2R_\mathrm{AGN} \sqrt{\frac r{\mathrm{G} M_\mathrm{SMBH}}}\nonumber \\
&=4\;\mathrm{hours} \; M_6^{\frac12} \left( \frac {R_\mathrm{AGN}}{10\;R_\mathrm g} \right) \left(\frac r {1\;\mathrm{mpc}} \right)^{\frac12}.
\end {align}
The transit duration is of the order of hours for massive main sequence stars closest to the AGN, and weeks for the closest giants further out in the circumnuclear star cluster (see Table~\ref{tab:transit} below).

\begin {figure}
\includegraphics*[width=86mm]{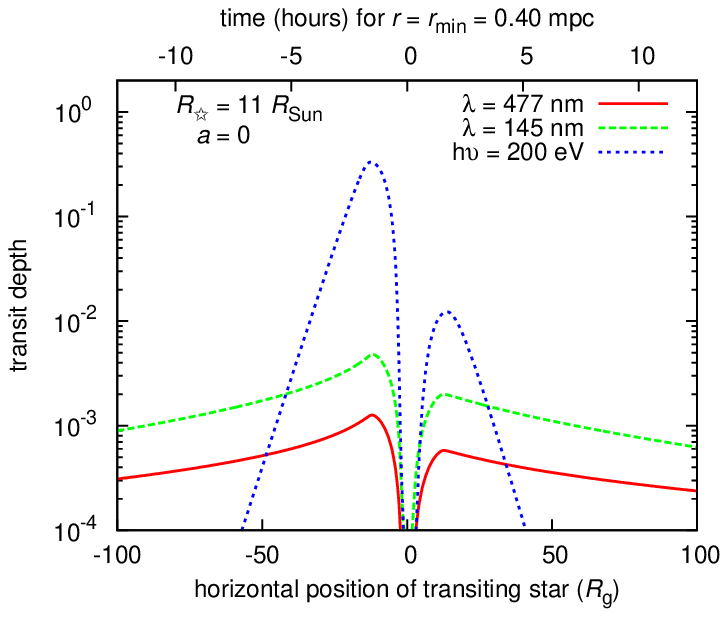}
\includegraphics*[width=86mm]{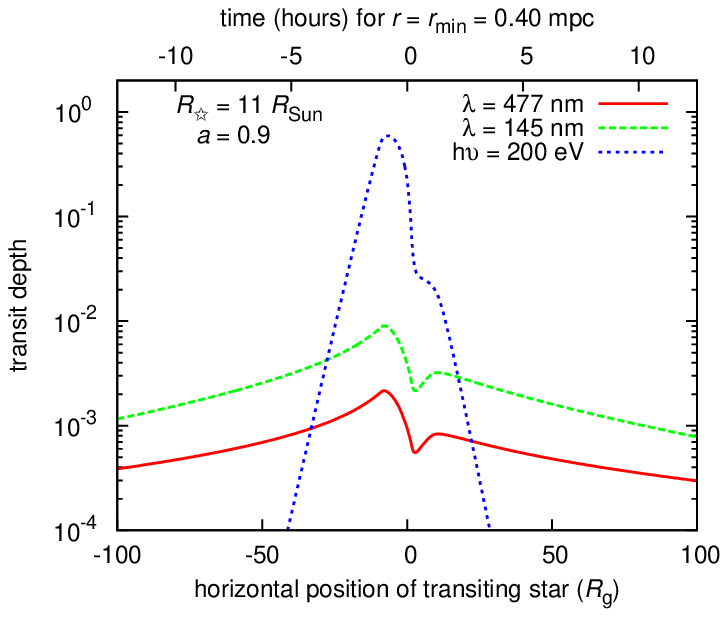}
\includegraphics*[width=86mm]{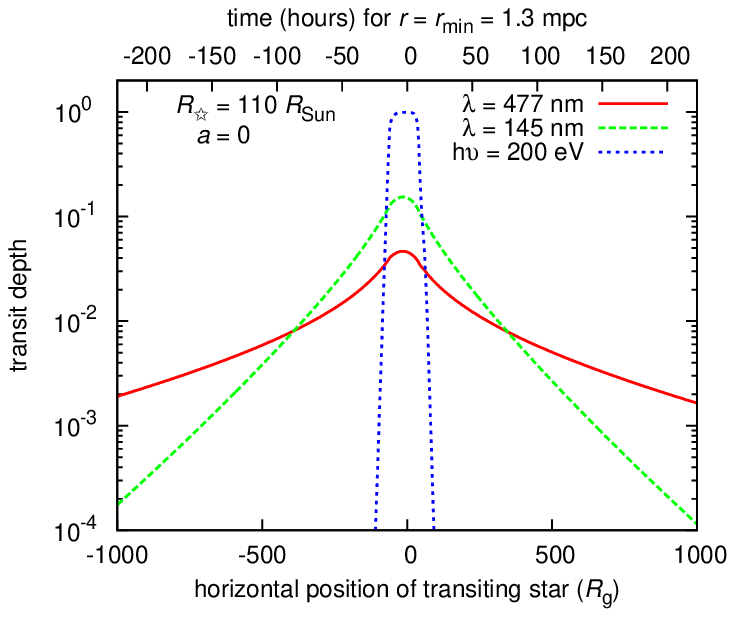}
\caption {Lightcurves of a transit: transit depth on logarithmic scale as a function of the horizontal position of the transiting star in front of the AGN in units of $R_\mathrm g$ (bottom axis), or time in case $r=5\;\mathrm{mpc}$ (top axis). Stellar radius is $R_\star = 11\;R_\sun = 5.2\;R_\mathrm g$ (top and middle panels) and $110\;R_\sun = 52\;R_\mathrm g$ (bottom panel). The SMBH spin is $a=0$ (top and bottom panels) and 0.9 (middle panel). The BH silhouette is larger than the star in the top panel only.}
\label {fig:lightcurve}
\end {figure}

Figure~\ref{fig:lightcurve} shows multiband transit lightcurves of an accretion disk due to stars of radius $R_\star = 11$ (top and middle panel) and $110\;R_\sun$ (bottom panel). The horizontal axis shows the horizontal position of the transiting star in the image plane of Figures~\ref{fig:transitmapostar}--\ref{fig:transitmapspin}. Physical distance in the projection plane is displayed in $R_\mathrm g$ units on the bottom axis, and the top axis shows time in hours for a star with minimum orbital radius as given in Table~\ref{tab:rlimits}. Note that further stars will cause a similar lightcurve, only scaled in time. The dimensionless black hole spin is zero on the top and bottom panels and $0.9$ in the middle panel. The projected stellar trajectory is horizontal with $y_\mathrm{proj}=1.5\;R_\mathrm g$, which approximately passes through the hottest part of the accretion disk. The black hole is behind the origin. The curves show the transit depth for the same three frequencies as the transit depth maps in Figures~\ref{fig:transitmapostar}--\ref{fig:transitmapgiant}. Note that these graphs are different from the usual planetary transit lightcurves showing flux, but plotting transit depth is more convenient when using a logarithmic scale. A larger value corresponds to a deeper transit, that is, a larger photometric signature. Figure~\ref{fig:lightcurve} is consistent with the expectations on the frequency dependence described above: at higher frequencies, the AGN is more compact, implying shorter and deeper transits. The sharp features near the center of the transit on the top and middle panels are due to the inner edge of the disk at the ISCO, and the left-right asymmetry is mostly due to Doppler shift of the approaching and receding parts of the disk. However, a transiting giant filters out this spatial feature due to its size, as seen on the bottom panel. The black hole silhouette is clearly visible in the top panel where the ISCO radius is larger than the transiting object. Comparing the top two panels, we infer that if the mass of the SMBH is known, observing a transit light curve allows us to put an upper limit on the spin.

We conclude that for an accretion disk around a $10^6\;M_\sun$ SMBH, $\sim0.1\%$ photometric accuracy is required in the optical, $\sim1\%$ accuracy in the extreme ultraviolet, and $\sim10\%$ accuracy in soft X-ray to detect a transit due to a 11 $R_\sun$ star. In order to detect a transit due to a 110 $R_\sun$ giant, $\sim1\%$ accuracy is sufficient in the optical, and $\sim10\%$ accuracy in extreme ultraviolet to X-ray (see Section~\ref{sec:agndensity} for a comparison with intrinsic variability).

For comparison, we ran simulations for two different SMBH masses: $10^5$ and $10^7\;M_\sun$. Table~\ref {tab:transit} shows the maximum possible transit depth in each case for different sizes of transiting stars. The transit depth for a fixed stellar size decreases with the increasing spatial extent of the accretion disk around black holes with increasing mass, making the transit feature more prominent for smaller BH masses. However, AGNs with less massive SMBHs are also intrinsically fainter, and $10^5\;M_\sun$ AGNs also have a smaller local number density which makes low mass AGN transit observations more challenging (see Section~\ref {sec:observability} for a discussion).

\section {Transit rates}
\label {sec:transits}

In this section, we estimate the expected rates of stellar transits in AGNs. Recall that $R_\mathrm{AGN}$ denotes the radius of the circular area that the center of the star has to transit in the projection plane for a given transit depth. Let $P_\mathrm{inst}(r)$ denote the probability of a single star on a circular orbit with radius $r$ transiting this circular area at a given instance, and let $P_\mathrm{ever}(r)$ denote the probability that the orbit is aligned so that the star transits this circular area at some point during its orbit. By geometric arguments, these probabilities are
\begin {eqnarray}
P_\mathrm{inst}(r) &=& \frac {\pi R_\mathrm{AGN}^2}{4\pi r^2} = \frac14 \frac {R_\mathrm{AGN}^2}{r^2}\,, \\
P_\mathrm{ever}(r) &=& \frac {2 R_\mathrm{AGN} \times 2\pi r} {4\pi r^2} = \frac {R_\mathrm{AGN}}r\,.
\end {eqnarray}

For a single AGN,
the expected value of the number of transits at a given instance, $N_\mathrm{inst}$,
the expected value of the transit rate (number of transits observed per time for asymptotically long observations), $\dot N$,
and the expected value of the number of stars on orbits such that they ever transit, $N_\mathrm{ever}$,
can be calculated by integrating for the entire stellar population:
\begin {align}
\label{eq:ninstpar}
N_\mathrm{inst} &= \int_{r_\mathrm{min}}^{r_\mathrm {max}} 4\pi r^2 n(r) P_\mathrm{inst}(r) \mathrm dr\,, \\
\dot N          &= \int_{r_\mathrm{min}}^{r_\mathrm {max}} 4\pi r^2 n(r) \frac {P_\mathrm{ever}(r)}{T_\mathrm{orb} (r)} \mathrm dr\,, \\
\label{eq:neverpar}
N_\mathrm{ever} &= \int_{r_\mathrm{min}}^{r_\mathrm {max}} 4\pi r^2 n(r) P_\mathrm{ever}(r) \mathrm dr\,,
\end {align}
where $T_\mathrm{orb}$ is the Keplerian orbital period.

To interpret these probability indicators, we have to understand the relationship between the expected value $\dot N$ of the transit rate and the expected value $N_\mathrm{ever}$ of the number of stars that ever transit a given AGN. If we monitor a single target for which $N_\mathrm{ever}\ll1$, then the actual transit rate is zero with probability $1-N_\mathrm{ever}$ and $\dot N/N_\mathrm{ever}$ with a small probability $N_\mathrm{ever}$. (The probability of multiple ever transiting stars in a single AGN is negligible in this case.) However, when monitoring a large number $n \gg 1/{N_\mathrm{ever}}$ of identical targets, these effects average out: the observed total transit rate is $\approx n\dot N$ and there are $\approx nN_\mathrm{ever}$ stars causing transits among all targets in total.

We substitute Equations~(\ref{eq:radiusofinfluence}--\ref{eq:density}) into Equations~(\ref{eq:ninstpar}--\ref{eq:neverpar}), and use $1<\alpha<3$ and $r_\mathrm{min}\ll r_\mathrm{max}$ (justified by Table~\ref{tab:rlimits}) to obtain
\begin {align}
\label {eq:ninstnum}
N_\mathrm{inst} &= 3.8\times10^{-6} \; 1000^{\alpha-2} \, b \frac{3-\alpha}{\alpha-1} M_6^{3 - \gamma (3-\alpha)}\;\times \\
\nonumber
&\hspace{-10mm}\times\left(\frac{M_\star}{30\;M_\sun}\right)^{-1} \left(\frac{R_\mathrm{AGN}}{10\;R_\mathrm g}\right)^2\left(\frac{r_0}{1\;\mathrm{pc}}\right)^{\alpha-3}
\left( \frac {r_\mathrm{min}}{1\;\mathrm{mpc}} \right) ^{1-\alpha} \\
\dot N &= \frac1{93\;\mathrm{yr}} \; 1000^{\alpha-2} \, b \frac{3-\alpha}{\alpha-\frac12} M_6^{\frac52-\gamma(3-\alpha)}\;\times \\
\nonumber
&\hspace{-10mm}\times\left(\frac{M_\star}{30\;M_\sun}\right)^{-1} \left(\frac{R_\mathrm{AGN}}{10\;R_\mathrm g}\right)  \left(\frac{r_0}{1\;\mathrm{pc}}\right)^{\alpha-3}
\left( \frac {r_\mathrm{min}}{1\;\mathrm{mpc}} \right) ^{\frac12-\alpha} \\
N_\mathrm{ever}^{\alpha>2} &= 0.032 \times 1000^{\alpha-2} \, b \frac{3-\alpha}{\alpha-2} M_6^{2-\gamma(3-\alpha)}\;\times \\
\nonumber
&\hspace{-10mm}\times\left(\frac{M_\star}{30\;M_\sun}\right)^{-1} \left(\frac{R_\mathrm{AGN}}{10\;R_\mathrm g}\right)  \left(\frac{r_0}{1\;\mathrm{pc}}\right)^{\alpha-3}
\left( \frac {r_\mathrm{min}} {1\;\mathrm{mpc}} \right) ^{2-\alpha} \\
N_\mathrm{ever}^{\alpha=2} &= 0.032 \times b (3-\alpha) M_6^{2-\gamma(3-\alpha)}\;\times \\
\nonumber
&\hspace{-10mm}\times\left(\frac{M_\star}{30\;M_\sun}\right)^{-1} \left(\frac{R_\mathrm{AGN}}{10\;R_\mathrm g}\right)  \left(\frac{r_0}{1\;\mathrm{pc}}\right)^{-1}
\ln \frac{r_\mathrm{max}}{r_\mathrm{min}} \\
\label {eq:neverless2}
N_\mathrm{ever}^{\alpha<2} &= 0.032 \times 1000^{\alpha-2} \, b \frac{3-\alpha}{2-\alpha} M_6^{2-\gamma(3-\alpha)}\;\times \\
\nonumber
&\hspace{-10mm}\times\left(\frac{M_\star}{30\;M_\sun}\right)^{-1} \left(\frac{R_\mathrm{AGN}}{10\;R_\mathrm g}\right)  \left(\frac{r_0}{1\;\mathrm{pc}}\right)^{\alpha-3}
\left( \frac {r_\mathrm{max}} {1\;\mathrm{mpc}} \right) ^{2-\alpha}.
\end {align}
Here $N_\mathrm{ever}^{\alpha>2}$, $N_\mathrm{ever}^{\alpha=2}$, and $N_\mathrm{ever}^{\alpha<2}$ denote the value of $N_\mathrm{ever}$ for the corresponding population density exponents. Stars on close orbits transit more frequently, and they dominate $N_{\mathrm{inst}}$ and $\dot N$. However, $N_\mathrm{ever}$ is dominated by stars on close or wide orbits for $\alpha>2$ and $\alpha<2$, respectively. This is determined by the exponent of $r$ in the integrand of Equations~(\ref{eq:ninstpar}--\ref{eq:neverpar}).

\begin{deluxetable*}{ccccc|ccccccc}
\tabletypesize{\scriptsize}
\tablecaption{Transit probabilities and duration for different bands and transit depths for a single AGN \label{tab:transit}}
\tablehead{&&&& maximum \\ $M_\mathrm{SMBH}$ & $a$ & model & $\lambda$ & transit & transit & $R_\mathrm{AGN}$ & $N_\mathrm {inst}$ & $1/{\dot N}$ & $N_\mathrm{ever}$ & $\Delta t\left(r_\mathrm{min}\right)$ & $\Delta t\left(r_\mathrm{max}\right)$ \\
$(M_\sun)$ &&& (nm) & depth\tablenotemark{a} & depth\tablenotemark{b} & $(R_\mathrm g)$ && (yr) && (h) & (h) }
\startdata
$10^5$ & 0   & O star    & 477 & 0.038    & $10^{-2}$ & 180 & $3.8\times10^{-6}$ &    280 & 0.009 &     89 &     750 \\
       &     &           & 145 & 0.076    & $10^{-2}$ & 235 & $6.5\times10^{-6}$ &    210 & 0.012 &    120 &     980 \\
       &     &           & 6.2 & $\sim1$  & $\sim1$   &  45 & $2.4\times10^{-7}$ & 1\,100 & 0.002 &     22 &     190 \\
$10^5$ & 0.9 & O star    & 477 & 0.049    & $10^{-2}$ & 200 & $4.7\times10^{-6}$ &    250 & 0.010 &     99 &     830 \\
       &     &           & 145 & 0.12     & $10^{-2}$ & 240 & $6.8\times10^{-6}$ &    210 & 0.012 &    120 &    1000 \\
       &     &           & 6.2 & $\sim1$  & $\sim1$   &  40 & $1.9\times10^{-7}$ & 1\,300 & 0.002 &     20 &     170 \\
$10^5$ & 0   & red giant & 477 & 0.63     & $10^{-1}$ & 900 & $1.2\times10^{-5}$ & 1\,000 & 0.11  &    830 & 20\,000 \\
       &     &           & 145 & 0.78     & $10^{-1}$ & 750 & $8.6\times10^{-6}$ & 1\,200 & 0.091 &    690 & 17\,000 \\
       &     &           & 6.2 & 0.74     & $10^{-1}$ & 530 & $4.3\times10^{-6}$ & 1\,700 & 0.064 &    490 & 12\,000 \\
$10^5$ & 0.9 & red giant & 477 & 0.68     & $10^{-1}$ & 850 & $1.1\times10^{-5}$ & 1\,100 & 0.10  &    780 & 19\,000 \\
       &     &           & 145 & 0.88     & $10^{-1}$ & 750 & $8.6\times10^{-6}$ & 1\,200 & 0.091 &    690 & 17\,000 \\
       &     &           & 6.2 & $\sim1$  & $10^{-1}$ & 500 & $3.8\times10^{-6}$ & 1\,800 & 0.060 &    460 & 11\,000 \\
$10^6$ & 0   & O star    & 477 & 0.0012   & $10^{-3}$ &   6 & $5.6\times10^{-7}$ &    320 & 0.010 &    1.5 &     7.9 \\
       &     &           & 145 & 0.0047   & $10^{-3}$ &  50 & $3.9\times10^{-5}$ &     39 & 0.083 &     13 &      66 \\
       &     &           & 6.2 & 0.33     & $10^{-1}$ &   9 & $1.2\times10^{-6}$ &    220 & 0.015 &    2.3 &      12 \\
$10^6$ & 0.9 & O star    & 477 & 0.0021   & $10^{-3}$ &  14 & $3.0\times10^{-6}$ &    140 & 0.023 &    3.5 &      18 \\
       &     &           & 145 & 0.0089   & $10^{-3}$ &  70 & $7.6\times10^{-5}$ &     28 & 0.13  &     18 &      92 \\
       &     &           & 6.2 & 0.56     & $10^{-1}$ &   8 & $9.9\times10^{-7}$ &    240 & 0.012 &    2.0 &      11 \\
$10^6$ & 0   & red giant & 477 & 0.047    & $10^{-2}$ & 220 & $8.4\times10^{-5}$ &    170 & 0.78  &     98 &  2\,900 \\
       &     &           & 145 & 0.15     & $10^{-1}$ &  55 & $5.3\times10^{-6}$ &    700 & 0.20  &     25 &     710 \\
       &     &           & 6.2 & $\sim1$  & $\sim1$   &  40 & $2.8\times10^{-6}$ &    960 & 0.14  &     18 &     520 \\
$10^6$ & 0.9 & red giant & 477 & 0.064    & $10^{-2}$ & 220 & $8.4\times10^{-5}$ &    170 & 0.78  &     98 &  2\,900 \\
       &     &           & 145 & 0.23     & $10^{-1}$ &  75 & $9.8\times10^{-6}$ &    510 & 0.27  &     34 &     980 \\
       &     &           & 6.2 & $\sim1$  & $\sim1$   &  40 & $2.8\times10^{-6}$ &    960 & 0.14  &     18 &     520 \\
$10^7$ & 0   & O star    & 477 & 0.000027 & $10^{-5}$ &  17 & $5.8\times10^{-4}$ &    4.4 & 0.95  &    2.2 &     7.1 \\
       &     &           & 145 & 0.00016  & $10^{-4}$ &   5 & $5.1\times10^{-5}$ &     15 & 0.28  &   0.64 &     2.1 \\
       &     &           & 6.2 & 0.0096   & $10^{-3}$ &   8 & $1.3\times10^{-4}$ &      9 & 0.45  &    1.0 &     3.3 \\
$10^7$ & 0.9 & O star    & 477 & 0.000055 & $10^{-5}$ &  35 & $2.5\times10^{-3}$ &    2.2 & 2.0   &    4.5 &      15 \\
       &     &           & 145 & 0.00032  & $10^{-4}$ &  12 & $2.9\times10^{-4}$ &    6.3 & 0.67  &    1.5 &     5.0 \\
       &     &           & 6.2 & 0.025    & $10^{-3}$ &   6 & $7.3\times10^{-5}$ &     13 & 0.34  &   0.77 &     2.5 \\
$10^7$ & 0   & red giant & 477 & 0.0020   & $10^{-3}$ &  17 & $5.8\times10^{-5}$ &     96 & 1.8   &    3.7 &     130 \\
       &     &           & 145 & 0.011    & $10^{-2}$ &   3 & $1.8\times10^{-6}$ &    540 & 0.32  &   0.65 &      23 \\
       &     &           & 6.2 & 0.55     & $10^{-1}$ &  12 & $2.9\times10^{-5}$ &    140 & 1.3   &    2.6 &      91 \\
$10^7$ & 0.9 & red giant & 477 & 0.0036   & $10^{-3}$ &  35 & $2.4\times10^{-4}$ &     47 & 3.7   &    7.6 &     270 \\
       &     &           & 145 & 0.021    & $10^{-2}$ &  12 & $2.9\times10^{-5}$ &    140 & 1.3   &    2.6 &      91 \\
       &     &           & 6.2 & 0.83     & $10^{-1}$ &   8 & $1.3\times10^{-5}$ &    200 & 0.84  &    1.7 &      61
\enddata
\tablenotetext{a}{The maximum transit depth is determined by model parameters: SMBH mass and spin, circumnuclear stellar population, and observing wavelength.}
\tablenotetext{b}{This transit depth is chosen to be less than the maximum transit depth. Subsequent columns show values dependent on this parameter. The same transit depth is chosen for corresponding $a=0$ and $a=0.9$ cases to demonstrate how little probability estimates depend on spin.}
\end{deluxetable*}

Let us substitute specific values to make numerical predictions. For concreteness, we study the cases of AGNs with mass $10^5$, $10^6$, and $10^7\;M_{\sun}$, and spin $a=0$ and $0.9$. Table~\ref {tab:transit} presents these examples. Each line of this table is generated by selecting a SMBH mass and spin, a stellar population model, and an observing wavelength. We determine the maximum possible transit depth by running our simulation with these input parameters. Then we choose a lower value as transit depth threshold, and use the calculated transit depth map (similar to Figures~\ref{fig:transitmapostar}--\ref{fig:transitmapgiant}) to identify the contour enclosing the area that a star has to transit to produce a lightcurve signature of at least this depth. We approximate this region by an ellipse, and take $R_\mathrm{AGN}$ to be the radius of a circular disk of the same area. These values can be substituted into Equations (\ref {eq:Deltat}) and (\ref{eq:ninstnum}--\ref{eq:neverless2}) to calculate the transit duration and to obtain probability estimates corresponding to the chosen transit depth treshold.

Table~\ref {tab:transit} shows that one cannot reasonably expect to detect transits when observing only a single or even a fistful of AGNs: typical transit rate is one every few hundred years (except for the very shallow transits for $M_\mathrm{SMBH}=10^7\;M_\sun$ which may happen more frequently than one in 10 years). However, these rate estimates are sensitive to poorly constrained parameters such as $r_\mathrm{min}$, $b$, and the number density exponent $\alpha$. Indeed, if we neglected the limits imposed by stellar collisions and set $r_\mathrm{min}$ to the tidal disruption radius, $r_\mathrm{min}$ would decrease by a factor of $\sim30$ or $\sim4$ (see Table~\ref{tab:rlimits}), and the instantaneous transit rate would increase a factor of $\sim200$ or $\sim3$, for O stars and giants, respectively. The event rates in a flattened, edge-on oriented star cluster may also be much larger \citep{2008ApJ...687..997S,2009ApJ...693.1959S}.

\section {Prospects for AGN transit observations}
\label {sec:observability}

In the previous section we established that it is necessary to monitor a large number of AGNs to confidently detect transit events. This can be done either by a targeted survey, or by monitoring a large region on the sky. In this section, we calculate the number density of suitable AGNs in the local universe, then discuss the feasibility of observing AGN transits with specific instruments. Note that there are other instruments potentially able to detect a transit, and archival data of previous observations might already contain transits.

\subsection {AGN density and variability}
\label {sec:agndensity}

The transit lightcurves are sensitive to the assumptions on the AGN, e.g.~on its mass and Eddington ratio. Larger mass means larger luminosity in optical to soft X-ray (but the disk is cooler, therefore less luminous in hard X-ray). Larger mass also means more extended accretion disk, that is, shallower transits. To quantify these effects, we consider three decades of magnitude ranges, centered on $M_\mathrm{SMBH}=10^5$, $10^6$, and $10^7\;M_\sun$. We integrate the lognormal fit of the local active black hole mass function determined by \citet {2007ApJ...667..131G,2009ApJ...704.1743G} to estimate the number density of AGNs in each mass range. Our results are given in Table~\ref {tab:dens}.

\begin{deluxetable}{ccc}
\tabletypesize{\scriptsize}
\tablecaption{Local active black hole density in three decades \label{tab:dens}}
\tablehead{SMBH mass range & number density\tablenotemark{a} \\
& $\left(\mathrm{Mpc}^{-3}\right)$}
\startdata
$10^{4.5}\;M_\sun < M_\mathrm{SMBH} < 10^{5.5}\;M_\sun$  & $5\times10^{-7}$ \\
$10^{5.5}\;M_\sun < M_\mathrm{SMBH} < 10^{6.5}\;M_\sun$  & $7\times10^{-6}$ \\
$10^{6.5}\;M_\sun < M_\mathrm{SMBH} < 10^{7.5}\;M_\sun$  & $1\times10^{-5}$
\enddata
\tablenotetext{a}{from the lognormal fit given by \citet {2007ApJ...667..131G,2009ApJ...704.1743G}}
\end{deluxetable}

For concreteness, we assume that all of these SMBHs are highly spinning and have a prograde accretion disk. However, Table~\ref {tab:transit} shows that probabilities depend weakly on the spin, therefore this assumption does not influence our estimates. (An exception is the case of X-ray observations, because accretion disk luminosity in this frequency depends strongly on spin, resulting in more potential targets, and thus more observable events down to a given luminosity limit.)

We assume the probabilities of the inclination $\cos\vartheta=0.5$ case for all AGNs (see Section \ref {sec:discussion} for a discussion of this assumption).

A transit can only be detected in the lightcurve if the AGN variability amplitude on the timescale of interest is small enough compared to the transit depth. \citet {2011A&A...525A..37M} investigate a sample of over 9000 quasars in the SDSS sample between $z=0.2$ and 3, and find that 93\%, 97\%, 93\%, 87\%, and 37\% of them are variable in the $u$, $g$, $r$, $i$, and $z$ band, respectively. However, the variability is dominated by timescales of months to years, much larger than the AGN transit
timescale of hours. A useful measure of the time dependence of the AGN variability is the structure function (SF), which essentially measures the RMS magnitude difference as a function of time lag $\tau$ between magnitude measurements. Based on the SDSS sample, \citet{2012ApJ...753..106M} find that for $\tau\lesssim 3$ year, $\mathrm{SF} = 0.02 \;\mathrm{mag}\;(\tau/10\;\mathrm{day})^{0.44}$. These observations are consistent with the damped random walk model of AGN variability \citep{2009ApJ...698..895K}.
Substituting a timescale of 1 hour, 1 day, and 1 week into this relation yields an average AGN variability of $0.0018$, $0.007$, and $0.017$ magnitudes in the optical $g$ band, which corresponds to a variability of $0.17\%$, $0.7\%$, and $1.6\%$ over these timescales, respectively. Comparing this to the AGN transit lightcurve on Figure~\ref{fig:lightcurve} for transiting giants shows that the transit can cause a larger change in luminosity than the average intrinsic optical AGN variability, provided that the observation cadence is at most a few days. Thus we conclude that red giant transits may be detectable even in typically variable AGN. However, AGN variability may be a limitation for detecting transiting O stars in the optical bands, where the transit depth is much smaller.

AGN variability is more significant on the transit timescale in X-rays \citep{2012arXiv1205.4255G,2012A&A...542A..83P}. However, in these bands, the transit may be a nearly complete eclipse (Figure~\ref{fig:lightcurve}), making them detectable regardless of variability. Indeed, transits of broad line clouds have already been detected in X-rays \citep{1998ApJ...501L..29M,2010A&A...517A..47M,2011MNRAS.410.1027R,2012ApJ...749L..31L}. Broad line clouds are expected to be much more densely distributed around AGNs then stars \citep{1997MNRAS.288.1015A,2006ApJ...636...83L}. However, their transit shapes are different from those of stars due to their cometary shape, with high column density heads followed by lower column tails \citep{2010A&A...517A..47M}. Some of these transiting clouds may represent the irradiated envelopes of circumnuclear ``bloated stars'' \citep{1980MNRAS.190..757E,1988MNRAS.233..601P} on very close orbits \citep{2012ApJ...749L..31L}.

\subsection {Ground-based optical instruments}
\label {sec:optical}

Pan-STARRS \citep {2002SPIE.4836..154K,2010SPIE.7733E..12K} is an optical and NIR survey project, consisting of four units with 1.8 m primary mirror diameter and a field of view of $7\;\mathrm{deg}^2$ each. The photometric precision is $\approx1$\% in most bands per $\lessapprox40$ s exposure. The first telescope has been operating in science mode since 2010. Once all four units are online, the system will survey the night sky once every week with a $5\sigma$ detection limit down to $r\approx24$. We use the conservative magnitude limit $g\approx23$ for 1\% photometric accuracy per single exposure, and calculate for ten years of operation of all four units.

An AGN with $M_\mathrm{SMBH}\sim 10^6\;M_\sun$, spin $a=0.9$, and Eddington ratio $0.3$ has $g=23$ magnitude if observed from a luminosity distance of $\approx330\;$Mpc. (For comparison, a similar AGN with $a=0$ would have the same $g$ magnitude from $390\;$Mpc.)
Using \texttt{cosmocalc.py}\footnote{by Tom Aldcroft, \url{http://cxc.harvard.edu/contrib/cosmocalc/}} to calculate the comoving volume up to this luminosity distance, and using the AGN number density in Table~\ref{tab:dens}, we find that there are approximately $600$ AGNs in the given mass range to this distance observable from a single geographic location. Main sequence O stars do not cause deep enough transits in the optical bands for $10^6\;M_\sun$ SMBHs, therefore we focus on transiting red giants. Assuming all targets host red giants, there will be $\sim35$ transits in ten years according to the transit rate $\dot{N} = 1/(170\;\mathrm{yr})$ given in Table~\ref {tab:transit}. When such an event occurs, the AGN is on the night sky with probability $\approx1/2$. The transit lasts longer than a week, therefore there will be at least one observation during the transit. Assuming $0.75$ of the dark hours have photometric conditions at the site, we estimate that the Pan-STARRS survey can detect $\sim10$ stellar transits in $\sim 10^6\;M_\sun$ AGNs. (If all targets had $a=0$, the expected number of transits would be $\sim20$.)

A similar calculation for highly spinning $10^5\;M_\sun$ AGNs gives an approximate 80 Mpc distance limit for a $g=23$ magnitude. However, in the catalog of \citet {2007ApJ...667..131G,2009ApJ...704.1743G}, there is only one active black hole with $M_\mathrm{SMBH} < 10^{5.5}\;M_\sun$ closer than this distance, therefore a Pan-STARRS detection of a transit event in the lightcurve of such an object is unlikely in the standard all-sky survey mode. Finally, there are approximately $80\,000$ AGNs with masses $\sim 10^7\;M_\sun$ within $1.65\;\mathrm{Gpc}$ corresponding to $g\approx23$ mag. We do not expect 1\% deep transits in this case. The transit rate at a $10^{-3}$ photometric level is $1/(47\;\mathrm{yr})$, therefore the Pan-STARRS survey would observe $\sim6\,000$ such events during 10 years of operation. Note however that detecting these transits may be prohibitively difficult due to systematics of ground-based observations, and also due to AGN variability.

Future ground-based optical surveys of larger collecting area might have an even better chance to detect transits of $10^6\;M_\sun$ active SMBHs. For example, LSST is a survey telescope with an equivalent primary mirror diameter of 6.68 m, and a field of view of $9.6\;\mathrm{deg}^2$, currently in design and development phase \citep {2009arXiv0912.0201L}. It is planned to have an observing strategy similar to that of Pan-STARRS, but with 3.4 times the collecting area of the four Pan-STARRS units in total, it can survey an approximately $3.4^{3/2}\approx 6$ times larger volume, increasing the transit detection expectation accordingly. Therefore we estimate that LSST may detect $\sim100$ transits per decade for $M_\mathrm{SMBH}\sim 10^6\;M_\sun$. Also, since LSST is planned to have more frequent visits than Pan-STARRS, the intrinsic variability will have a smaller amplitude between subsequent observations.

Since many survey telescopes devote a certain fraction of their time to deeper surveys of smaller areas, it is worth estimating how this changes the expected number of transit detections. Recall that the four Pan-STARRS units will be able to scan the entire night sky approximately once a week in five filters with $\approx40$ second exposures. Now assume a single Pan-STARRS class instrument spends all available time on surveying an $n$ times smaller area in $g$ filter only, with $m$ visits per week to fight AGN variability, thus exposing $n/m$ times longer. As long as the observations are photon noise limited and cosmological effects are negligible, this strategy increases the surveyed volume, and also the expected number of transit detections, by a factor of $n^{1/2}m^{-3/2}$. For example, daily visits ($m=7$) of an $n\approx1300$ smaller area (a few fields) would mean two hour exposures per pointing, and would double the expected transit detections around $M_\mathrm{SMBH}\sim 10^6\;M_\sun$ AGNs to $\sim2$--4 per year, with increased robustness against intrinsic variability.

\subsection {Kepler observations}
\label {sec:kepler}

The Kepler satellite carries out almost continuous photometric measurements in the bandpass of 420 to 900 nm. It achieves $\approx10^{-4}$ photometric accuracy per 30 minute exposure on $Kp=13$ dwarf stars \citep {2010ApJ...713L..79K}.

\citet {2011ApJ...743L..12M} report observations of four AGNs, and \citet {2012arXiv1203.1942E} identify 13 more in the Kepler field. However, these quasars are at redshifts between $z=0.028$ and $0.625$, therefore it is not likely that either of them has low enough SMBH mass that transits due to O stars or red giants could be observed by Kepler.

\subsection {Space X-ray instruments}
\label {sec:roentgen}

The Chandra X-ray observatory has been operating in orbit since 1999. Its relevant instrument is the High Resolution Camera (HRC) has a 31 arcmin by 31 arcmin field of view, sensitive from 0.1 keV to 10 keV \citep {2002PASP..114....1W}.

The X-ray Multi-Mirror Mission (XMM) Newton satellite is another proposal instrument, operating since 2000.
It is equipped with
three imaging cameras sensitive from 0.2 keV to 12 keV \citep {2001A&A...365L..27T,2001A&A...365L..18S},
with approximately $700\;\mathrm{arcmin}^2$ field of view each.

The Spektr-RG satellite is scheduled to launch in 2013 at earliest. Its eROSITA X-ray telescope system has approximately twice the effective area of a single instrument on XMM-Newton below 2 keV \citep {2010SPIE.7732E..23P}. This observatory is planned to carry out a survey consisting of eight full scans of the sky in four years, with a mean total exposure time of 2 ks for each region \citep {2011xru..conf..192B}.

The Extreme Physics Explorer \citep {2011arXiv1112.1327G} is a mission concept designed specifically to study accretion disks around SMBHs. It is proposed to have more than an order of magnitude larger effective area than current missions, thus capable of observing targets farther away.

As an example, we estimate the expected number of transits detected during a single 100 ks observing campaign with the HRC-I instrument on Chandra. We consider observations in the energy range 0.1 keV to 0.4 keV, logarithmically centered on the energy 0.2 keV for which we calculated transit probabilities. Based on Section~\ref{sec:spectra}, the integrated luminosity in this range is $9\times10^{41}\;{\mathrm{erg}}/{\mathrm{s}}$, $8\times10^{42}\;{\mathrm{erg}}/{\mathrm{s}}$, and $6\times10^{43}\;{\mathrm{erg}}/{\mathrm{s}}$, and the photon index is $-0.2$, $0.4$, and $1.6$, for AGNs with $a=0.9$ and mass $10^5$, $10^6$, and $10^7\;M_\sun$, respectively. (Accretion disks around non-spinning black holes are up to a factor of four less luminous in this energy range.)

Let us divide the observation time into ten bins of 10 ks each for temporal resolution. If we want the relative photon noise to be 0.1 in each bin, we need a count rate of at least $0.01$ photons per second. According to the PIMMS count rate calculator\footnote{\url {http://cxc.harvard.edu/toolkit/pimms.jsp}}, this is the case up to a luminosity distance of $\approx290$, $890$, and $2\,300\;$Mpc for SMBH masses of $10^5$, $10^6$, and $10^7\;M_\sun$, respectively. This corresponds to a redshift $z\approx0.06$, $0.2$, and $0.4$, which is small enough to neglect in terms of flux change for a simple estimate. There are $\approx40$, $\approx12\,000$, and $\approx180\,000$ AGNs in the corresponding mass ranges, out to this luminosity distance, out of which $\approx0.0003$, $\approx0.08$, and $\approx1$ fall in the Chandra field of view on average. Therefore we can only observe a single or at most a few targets in one pointing. Since the probability of detecting a transit when observing a single target for this much time is negligible based on the transit rate value of Table~\ref {tab:transit}, such a short campaign is not suitable for stellar AGN transit discovery.

Now let us consider the planned eROSITA survey. Suppose the 2 ks exposure time is evenly distributed among eight visits of 250 s on each object. We have the same 0.1 relative photon noise per visit out to approximately 200\;Mpc, 700\;Mpc, and 2\;Gpc, for AGNs with $10^5$, $10^6$, and $10^7\;M_\sun$, respectively. On the entire sky, this means approximately 20, 6\,000, and 100\,000 AGNs. For such short observations, the expected number of transits sampled is the product of the number of AGNs, the number of visits per AGN, and the instantaneous probability of a transit, $N_\mathrm{inst}$. Based on the values in Table~\ref {tab:transit}, if we assume O stars around each target, we expect $3\times10^{-5}$ almost total eclipses for AGNs with mass $\sim10^5\;M_\sun$, and $0.05$ transits of depth $0.1$ for AGNs with $10^6\;M_\sun$. In case of SMBH mass around $10^7\;M_\sun$, transits due to O stars do not reach the depth of $0.1$, therefore they are not likely to be detected in the inherent variability. If we assume a $b=0.01$ mass fraction of giant stars in a typical AGN, we expect approximately $\sim10^{-3}$, $0.1$, and $10$ transits of depth $\sim1$, $\sim1$, and $0.1$ detected for SMBH masses $\sim10^5$, $10^6$, and $10^7\;M_\sun$, respectively.

We conclude that discovering stellar transits in X-ray with proposal instruments is unlikely, whereas it may be possible with survey instruments like eROSITA for AGNs with $10^6$--$10^7\;M_\sun$.

\section {Discussion}
\label {sec:discussion}

In this section we highlight the uncertain points in our argument, not only to be able to properly interpret our predictions, but also to understand the implications of future AGN transit detections or non-detections.

We based our stellar models in galactic nuclei on observations of the center of the Galaxy in \textsection\ref {sec:cusps}, assuming that this is a good representation of circumnuclear star clusters in active galaxies, and on theoretical dynamical models. We set up simplified stellar population models to estimate transit rates. Our estimates of transit probabilities depend strongly on the assumed value of the stellar distribution exponent $\alpha$: the $1000^{\alpha-2}$ terms in Equations (\ref {eq:ninstnum})--(\ref {eq:neverless2}) are due to the three order of magnitude difference in the distance scales of $r_\mathrm{min}$ and $r_0$. As a consequence, steeper stellar distributions feature larger transit probabilities, consistent with the statement that most probabilistic measures are dominated by closeby stars for moderate values of $\alpha$.

\begin {figure}
\includegraphics*[width=86mm]{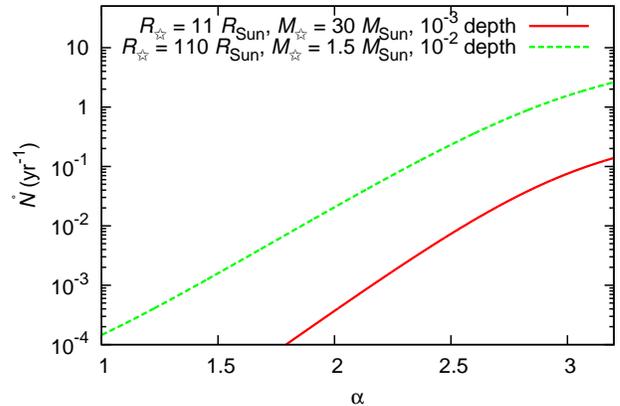}
\caption {Transit rate $\dot N$ for a single target AGN as a function of stellar density exponent $\alpha$, with $M_6=1$, $a=0.9$, and $b=0.01$. Solid red curve corresponds to O stars at a $g$-band transit depth threshold of $10^{-3}$, dashed green curve corresponds to red giants or bloated stars at a $g$-band transit depth threshold of $10^{-2}$.}
\label {fig:ndotofalpha}
\end {figure}

Evaluated at $\alpha=3$, Equation (\ref {eq:density}) and thus Equations (\ref{eq:ninstnum}--\ref{eq:neverless2}) yield zero. This artifact originates in the assumption that the number of stars within $r_\mathrm{min}$ is negligible, which is not true for this value of $\alpha$. To study the case of $\alpha\approx3$, we recalculate $\dot N$ without this assumption:
\begin {align}
\dot N &= \frac b{0.093\;\mathrm{yr}} \frac{3-\alpha}{\alpha-\frac12} M_6^{\frac52}\left(\frac{M_\star}{30\;M_\sun}\right)^{-1} \times \\
\nonumber
&\times \frac{r_\mathrm{min}^{3-\alpha}}{r_\mathrm i^{3-\alpha}-r_\mathrm{min}^{3-\alpha}} \left( \frac {r_\mathrm{min}}{1\;\mathrm{mpc}} \right) ^{-\frac52}\left(\frac{R_\mathrm{AGN}}{10\;R_\mathrm g}\right).
\end {align}
Figure \ref {fig:ndotofalpha} displays this formula for $\dot N$ as a function of $\alpha$. We plot two cases: transit rate with a $g$-band threshold depth of $10^{-3}$ for O stars ($R_\mathrm{AGN}=14$), and with a $g$-band threshold depth of $10^{-2}$ for red giants or bloated stars with $R_\star=110\;R_\sun$, $M_\star=1.5\;M_\sun$ ($R_\mathrm{AGN}=220$). We fix $M_6=1$, $a=0.9$, and $b=0.01$ for both cases. The plots show us that $\dot N$ is a steeply increasing function of $\alpha$ up to and over the value 3, as expected. Comparing the plot to the values given in Table \ref {tab:transit} at $\alpha=2.5$ in the first case ($\dot N = 1/140\;\mathrm{yr}$) and at $\alpha=1.75$ in the second case ($\dot N=1/170\;\mathrm{yr}$) tells us that our approximation in Equation (\ref {eq:density}) for these values of $\alpha$ is valid.

Our method to determine the number of stars within the sphere of influence based on the $M$-$\sigma$ relation is oversimplified. The actual number of stars depends on formation, relaxation, and disruption rate and history. Furthermore, the presence of stellar remnants may dilute the stellar population within the sphere of influence for a fixed total mass, decreasing the mass fraction $b$ -- and therefore the number -- of stars that would potentially cause detectable transits.

At large orbital radii, we expect microlensing to dominate over the light obscuration due to the transiting star. We did not investigate this phenomenon in detail but conservatively ignored all transits due to stars on wide orbits where microlensing may become significant. As the transit rate is dominated by closeby stars, microlensing is not a limiting factor of transit observability after all. Also, we did not account for apsidal and Lense--Thirring precession, which may move stars in and out of transiting orbits.

AGN transits due to broad line clouds are expected to be much more frequent than the stellar transits calculated here \citep [see e.g.] [] {1998ApJ...501L..29M, 2009ApJ...695..781B, 2010A&A...517A..47M, 2007ApJ...659L.111R, 2009MNRAS.393L...1R, 2009ApJ...696..160R, 2011MNRAS.410.1027R,2011MNRAS.417..178R}. Transits of clouds may be somewhat different than stellar transits due to their cometary shape \citep{2010A&A...517A..47M}, and typically have a shorter duration due to their proximity to the SMBH and larger velocity (typically $\Delta t\sim 1\;\mathrm{hr}\;M_6$ if orbiting at $r_\mathrm{BLR}\sim 1000\;R_g$). 

Note that we have assumed that the transiting stars move on orbits much wider than the accretion disk. This assumption may be violated for stars on very close orbits ($r\lesssim 10^5 \; R_g = M_6 \; 4.8 \;\mathrm{mpc}$), where stars crossing the disk may get captured by hydrodynamic drag \citep {2001A&A...376..686K}.

We mentioned the possibility of supermassive stars possibly forming in AGN accretion disks, but do not have information on their occurrence rates. Furthermore, Wolf--Rayet stars exhibit strong stellar winds, which might form an optically thick region of radius $\sim100\;R_\sun$ \citep [Figure 5 in][and references therein] {2007ARA&A..45..177C}; OH/IR stars can form dust clouds much larger than the entire accretion disk \citep [e.g.][] {1987A&A...186..136B,2002A&A...384..585K}; and bloated stars with large irradiated envelopes \citep{1980MNRAS.190..757E,1988MNRAS.233..601P} may be present near the AGN. These objects can potentially cause much deeper transits or even eclipses.

In Section~\ref {sec:spectra}, we investigated the projected thermal radiation structure of an accretion disk with specific parameters, assuming radiatively efficient accretion, modelling radiative transfer in the accretion disk photosphere, and accounting for light propagation in the Kerr metric of the SMBH. We set the Eddington ratio to $0.25$, a value based on the analyses of \citet {2006ApJ...648..128K} and \citet {2011arXiv1111.3574S}. This value is consistent with the findings of \citet {2007ApJ...667..131G}. However, keep in mind that magnitude limited samples are biased towards larger Eddington ratios. Also note that the Eddington ratio can in fact vary by an order of magnitude in either direction \citep[e.g.][]{2009MNRAS.397..135K}. A larger accretion rate would increase the total luminosity and the peak frequency with little effect on the size of the accretion disk. Therefore such a disk would be much brighter in X-ray than one with lower Eddington ratio, while they would exhibit transits of similar depth.

We also assumed prograde disk alignment for spinning SMBHs, but noted that the transit probabilities do not depend strongly on spin. We fixed the inclination at $\cos\vartheta=0.5$, which is the mean value for an isotropic distribution, therefore the predicted transit probabilities are typical of all inclinations. However, note that larger inclination results in a thinner image of the accretion disk, therefore deeper transits. A disk with thickness $H/R\approx 0.05$ observed from a nearly edge-on orientation results in a $\sim10$ times deeper transit, leading to a $10^3$ times larger detectable volume for photon noise limited surveys. On the other hand, there are selection effects at both inclination extrema: a coplanar torus might obscure thermal emission from edge-on AGNs, whereas jets along the rotational axis might contaminate the lightcurve of face-on AGNs and prevent transits from being detected. These are likely to confine the inclination distribution of AGNs with observable thermal radiation closer to the average values we use in our model.

Intrinsic AGN variability will pose a challenge to identifying AGN transits. Fortunately, optical AGN variability is small on the timescales of days, and thus does not rule out the possibility of transit observations for giant stars. With Pan-STARRS observations, however, the variability level at typical observation cadence is comparable to the transit depth. Transits might not be detected if the weekly observations miss the deepest part.

Simultaneous multiband observation campaigns can help distinguish variability from transit signiture, as the transit lightcurve is predicted to have a different shape in different frequency bands. Also, transit lightcurves could be contaminated by the flux reflected by clouds surrounding the AGN. Multi-wavelength campaigns following AGN for these types of events may give information about the reflecting fraction, which could allow to constrain the covering fraction of material, the location and relative sizes of the reflecting regions. Future detectors could distinguish the reflected component using polarimetry.

\section {Conclusions}
\label {sec:conclusions}

In this paper, we presented simple estimates to study the prospects for detecting stellar AGN transits. We have shown that such observations would offer a novel possibility to image the accretion disk in distant AGN with unprecedented accuracy. For example, the black hole silhouette (i.e., the lack of emission within the ISCO) may be resolved using the lightcurve of an AGN transit due to an O-type main sequence star. These observations probe the accretion disk and the space-time geometry around black holes, and in particular, they are sensitive to the black hole spin. AGN stellar transit event rates offer information about the circumnuclear stellar cluster.

We predict that the Pan-STARRS survey could detect $10\sim20$ stellar transits, and LSST may detect $\sim100$ by repeated photometric observations of $\sim10^6\;M_\sun$ AGNs. We estimate that stellar transit detections in X-rays are not likely with individual campaigns on proposal instruments, but we expect a few possible detections in $\sim10^7\;M_\sun$ AGNs with X-ray surveys like eROSITA. Note that these rate estimates do not include transits by Compton thick clouds, which are observed to be common in X-rays. The transit rate corresponding to clouds could also be much more frequent in optical surveys.

However, these probability estimates are very sensitive to parameters which are based on theoretical arguments. One of these parameters is the inner radius cutoff of the nuclear stellar cluster, $r_\mathrm{min}$, which is set by stellar collisions. We have shown that if stellar collisions did not deplete the innermost regions, the less stringent limitation due to tidal disruption would imply $\sim200$ times larger transit rates for O stars. Such scenarios may be possible if the effective stellar size increases during close approach to the AGN, leading to large irradiated envelopes \citep[``bloated stars'', see][]{1980MNRAS.190..757E,1988MNRAS.233..601P}. AGN transit observations could constrain these parameters and refine circumnuclear stellar population models in distant galaxies.

\acknowledgements

The authors thank \'Akos Bogd\'an, Martin Elvis, Jeff McClintock, Barry McKernan, and Guido Risaliti for helpful discussions. BK acknowledges support from NASA through Einstein Fellowship PF9-00063 issued by the Chandra X-ray Observatory, operated by the SAO, on behalf of NASA under contract NAS8-03060.

\bibliography {paper}

\end {document}